\documentstyle{aipproc}
\input psfig
\newcommand{\pom}{I\!\! P}
\begin{document}
%
%
\vglue -1cm
\title{Pomeron Flux Renormalization:\\
A scaling Law in Diffraction\footnote{\normalsize
Presented at ``Diffractive Physics, 
LAFEX International School on High Energy Physics
(LISHEP-98), Rio de Janeiro, Brazil, 10-20 February 1998."}}
\author{Konstantin Goulianos}
\address{The Rockefeller University,
New York, NY 10021, U.S.A.}
\maketitle
\begin{abstract}
The pomeron flux renormalization hypothesis is reviewed and 
presented as a scaling law in diffraction. Predictions for soft 
and hard diffraction based on pomeron flux scaling 
are compared with experimental results. 
\end{abstract}
\section*{STANDARD POMERON FLUX}
The cross section for hadron dissociation on protons, 
$hp\rightarrow Xp$, at large $x_F\equiv p^*_{\|}/2\sqrt s$, where 
$p^*_{\|}$ is the $z$(beam)-component of the (leading) proton in the 
final state, is dominated by pomeron exchange~\cite{KG}.  
In Regge theory, the pomeron contribution is given by the 
triple-pomeron amplitude
\begin{equation}
\frac{d^2\sigma^{hp}_{sd}}{d\xi dt}=
\frac{{\beta_{\pom pp}^2(t)}}{16\pi}\;\xi^{1-2\alpha_{\pom}(t)}
\left[\beta_{\pom hh}(0)\,g(t)
\;\left(\frac{s'}{s_0}\right)^{\alpha_{\pom}(0)-1}\right]
\label{diffractive}
\end{equation}
where $\alpha_{\pom}(t)
=\alpha_{\pom}(0)+\alpha' t=(1+\epsilon)+\alpha' t$
is the pomeron       
trajectory, $\beta_{\pom pp}(t)$ is the coupling of the pomeron to the proton,
$g(t)$ is the triple-pomeron coupling, $s'=M_X^{2}$ 
is the $\pom-p$ center of
mass energy squared, $\xi \equiv 1-x_{F}=s'/s=M_X^2/s$ is the fraction of
the momentum of the proton carried by the pomeron, and $s_0$ is an energy
scale parameter not determined by the theory and 
usually set to 1~GeV$^2$ (the hadron mass scale).

The term in brackets in (\ref{diffractive})
has the form of the $\pom-p$ total cross section.
Thus, the process $hp\rightarrow Xp$ can be viewed as 
a flux of pomerons 
emitted by the proton interacting with the hadron $h$. 
The pomeron ``flux factor" is represented by   
\begin{equation}
f_{\pom /p}(\xi,t)\equiv \frac{{\beta_{\pom pp}^2(t)}}{16\pi}\;
\xi^{1-2\alpha_{\pom}(t)}
\equiv K\,\xi^{1-2\alpha_{\pom}(t)}\,F^2(t)
\label{flux}
\end{equation}
where $K\equiv \beta_{\pom pp}^2(0)/16\pi$ and $F(t)$ is the proton 
form factor.
Ingelman and Schlein (IS)~\cite{IS} 
proposed using this {\em standard} pomeron flux factor
in calculating hard single diffraction dissociation cross sections.
In such calculations, one assumes 
that the pomeron has a partonic structute 
and lets the partons of a $\pom$ coming from the proton 
interact with the partons in $h$.

There are two problems with the IS method in using the standard pomeron flux to 
calculate hard diffraction rates: 
\begin{enumerate}
\item The normalization of the standard flux depends on the energy 
scale $s_0$ through the total cross section equation, $\sigma_T^{pp}=
\beta_{\pom pp}^2(0)\cdot (s/s_0)^{\textstyle \epsilon}$. Since the scale $s_0$
is not determined by the theory, the value of $\beta_{\pom pp}(0)$,
and therefore that of the standard flux normalization, is arbitrary.
\item For any given value of the energy scale $s_0$, 
the diffractive cross section grows as $s^{2\textstyle \epsilon}$, overtaking 
at high energies the total cross section, which grows as 
$s^{\textstyle \epsilon}$,
in violation of unitarity~\cite{R}. 
\end{enumerate}
It is well known that the Regge theory  
$\sim s^{\textstyle \epsilon}$
dependence of $\sigma_{T}(s)$ itself violates the
unitarity based Froissart bound, which  states that the
total cross section cannot rise faster than  $\sim \ln^2 s$.
Unitarity is also violated by the $s$-dependence of the ratio
$\sigma_{el}/\sigma_T\sim s^{\textstyle \epsilon}$, 
which eventually exceeds the
black disc bound of one half ($\sigma_{el}\leq \frac{1}{2}\sigma_T$), 
as well as by the $s$-dependence of the $b=0$ value of the elastic scattering 
amplitude in impact parameter space, 
which has already reached a value close to the maximum 
allowed by unitarity at 
$\sqrt s=1.8$ TeV~\cite{CDF}.
However,
for both the elastic and total cross sections,
unitarization can be achieved by taking into account rescattering 
effects using the eikonal formalism~\cite{GLM,CMG}.
Attempts to introduce rescattering in the {\em diffractive} amplitude
by eikonalization~\cite{GLM} or by including cuts~\cite{K1,K2} have met
with moderate success. 
Through such efforts it has become clear that these
``shadowing" or 
``screening" corrections affect mainly the normalization of the
diffractive amplitude, leaving the form of the $M^2$ dependence almost
unchanged. This feature is clearly present in the data,
as demonstrated by the CDF Collaboration~\cite{CDF} in comparing 
their measured diffractive differential
$\bar p p$ cross sections at $\sqrt{s}=$546 and 1800 GeV  with $pp$
cross sections at $\sqrt{s}=20$ GeV.

Motivated by these
theoretical results  and by the trend observed in the data,
a phenomenological approach to unitarizing the diffractive amplitude
was proposed~\cite{R} based on ``renormalizing" the pomeron flux 
by requiring its integral over all available $\xi$ and $t$ to saturate at unity.
Such a normalization, which
corresponds to {\em a maximum} 
of one pomeron per proton, leads to interpreting the
pomeron flux as a probability density simply describing the $\xi$ and $t$
distributions of the exchanged pomeron in a diffractive process.

\section*{RENORMALIZED POMERON FLUX}
The renormalization of the  pomeron flux 
is based on a {\em hypothesis}, rather than on a 
calculation of unitarity corrections, and therefore can be 
stated as an axiom: 
\vglue 2ex
\noindent\fbox{
{\bf The pomeron flux integrated over all phase space saturates at unity.}
}
\vglue 2ex
Mathematically, the renormalized pomeron flux is given by 
\begin{equation}
f_N(\xi,t)=N^{-1}(\xi_{min})\cdot f_{\pom/p}(\xi,t)
\label{fluxN}
\end{equation}
The renormalization factor $N(\xi_{min})$ is the integral
of the flux
\begin{equation}
N(\xi_{min})=\displaystyle{\int_{\xi_{min}}^{0.1}}
\displaystyle{\int_{t=-\infty}^{t=0}}f_{\pom/p}(\xi,t)d\xi dt
\label{fluxI}
\end{equation}
where the upper limit of the integration over $\xi$ has been taken to be 
$\xi_{max}=0.1$ (the coherence limit~\cite{KG}).

The renormalized flux overcomes the two probems of the standard flux:
\begin{enumerate}
\item The normalization is no longer arbitrary, 
since the energy scale factor $s_0$ 
cancels out in dividing the standard flux by its integral. 
\item The diffractive cross section now grows as 
\begin{equation}
\sigma_{sd}\sim \int_\xi\int_t(s\xi)^{\textstyle \epsilon}f_N(\xi,t)d\xi dt
\sim s^{\textstyle \epsilon}\cdot
\left<\xi^{\textstyle \epsilon}\right>_{f_N}
\stackrel{s \rightarrow \infty}{\Rightarrow} \mbox{constant}
\label{sigmatot}
\end{equation}
and thus respects the unitarity bound.
\end{enumerate}
The renormalization factor is a function of $\xi_{min}$,
which is process dependent. Thus, conventional factorization breaks down.
The scaling of the pomeron flux to its integral can be viewed as 
\vglue 1ex
\centerline{\fbox{\bf A scaling Law in Diffraction}}
\vglue 1ex
which unitarizes the diffractive amplitude at the expense of factorization.

\section*{COMPARISON OF RENORMALIZED FLUX PREDICTIONS WITH DATA}
Predictions made using the renormalized pomeron flux have been 
compared with data for both soft~\cite{R,GM} and 
hard~\cite{R,KGDIS,PIC,Frascati,WJJ,DJJ,DPE} 
diffraction. In this section we summarize briefly the results of such 
comparisons. 

\subsection*{Soft Diffraction}
\begin{itemize}
\item The renormalized flux prediction of the $s$-dependence of 
the total $pp/\bar pp$ single diffractive cross section 
is in excellent agreement with the data~\cite{R}.
\item The differential cross section $d^2\sigma_{sd}/dM_X^2dt|_{t=0}$
for $pp/\bar pp$ is independent of $s$ and behaves as 
$\sim 1/(M_X^2)^{1+\textstyle \epsilon}$~\cite{GM}. This {\em scaling}
behavior, which holds over six orders of magnitude, 
is predicted by the 
renormalized flux~\cite{GM} and disagrees with the 
$\sim s^{2\textstyle\epsilon}$ standard flux expectation.
\end{itemize}
\subsection*{Hard Diffraction}
Hard diffraction has been studied at HERA and at $\bar pp$ colliders. 
In this section we discuss results on the pomeron structure 
obtained at HERA and at the Tevatron and compare measured 
diffractive production rates with predictions based on the standard and 
renormalized pomeron flux. 
  
\subsubsection*{Results from HERA}
At HERA, both the H1 and ZEUS Collaborations used deep inelastic 
scattering (DIS) to measure the ``diffractive 
structure function" of the proton, 
$F_2^{D(3)}(Q^2,\beta,\xi)$, where $\beta$ is 
the fraction of the momentum of the pomeron taken by the struck quark.
Both experiments found the form
\begin{equation}
F_2^{D(3)}(Q^2,\beta,\xi)=\frac{1}{\xi^{1+n}}\cdot A(Q^2,\beta)
\label{F2D3}
\end{equation}
in which the variable $\xi$ factorizes out into an expression
reminiscent of the pomeron flux factor. Therefore, it appeared reasonable to 
consider the term $A(Q^2,\beta)$ as being proportional to 
the pomeron structure function $F_2^{\pom}(Q^2,\beta)$. 
This term was found to be rather flat in $\beta$,
suggesting that the pomeron has a {\em hard} quark structure. 
For a fixed $\beta$, $A(Q^2,\beta)$ increases with $Q^2$. 
By interpreting the $Q^2$ dependence to be due to scaling violations, 
the H1 Collaboration extracted the gluon fraction of the pomeron 
using the DGLAP evolution equations in a QCD analysis of 
$F_2^{D(3)}(Q^2,\beta,\xi)$.
The ZEUS Collaboration determined the gluon fraction by combining 
information from diffractive DIS, which is sensitive mainly to the 
quark component of the pomeron, 
and diffractive dijet photoproduction, which is 
sensitive both to the quark and gluon contents.
Both experiments agree that the pomeron structure is hard and consists of 
gluons and quarks in a ratio of approximately $3\div 1$.
In both cases, the extracted 
gluon fraction does not depend on the pomeron flux normalization. 
\subsubsection*{Results from the Tevatron}
Both the CDF and D\O\, Collaborations have reported 
that the jet $E_T$ distributions from non-diffractive (ND), single 
diffractive (SD) and double pomeron exchange (DPE) dijet events 
have approximately the same shape~\cite{DJJ,DPE,DPED0}. 
Since in going from ND to SD or from SD to DPE 
a nucleon of momentum $p$ is replaced by a pomeron of momentum $p\xi$,  
the similarity of the $E_T$ spectra 
suggests that the pomeron structure must be harder than the 
structure of the nucleon by a factor of $\sim 1/\xi$.
Assuming a hard pomeron structure, the CDF Collaboration 
determined the gluon fraction of the pomeron to be $f_g=0.7\pm 0.2$ 
by comparing the measured rate of diffractive $W$ production, 
which is sensitive to the quark content of the pomeron, with the rate for 
diffractive dijet production, 
which depends on both the quark and gluon contents~\cite{WJJ}.
These results, which are independent of the pomeron flux normalization, 
agree with the results obtained at HERA.

For a hard pomeron structure with $f_g=0.7$ and $f_q=0.3$, 
the measured $W$ and dijet rates are smaller than the rates 
calculated using the standard flux by a factor $D=0.18\pm 0.04$.
This flux ``discrepancy" factor is consistent with the pomeron flux 
renormalization expectation~\cite{R,KGDIS}. 

The CDF Collaboration also measured the rate for DPE dijets and 
compared it with the rates for SD and ND dijets and with calculations 
using the standard pomeron flux~\cite{DPE}. 
To obtain the measured DPE/SD ratio, 
the standard flux in DPE must be multiplied by the factor $D$ {\em 
for both the proton and antiproton}. This result 
supports the hypothesis that the suppression factor, relative to the 
standard flux calculations, is associated with the flux, rather than 
with  ``screening corrections" as proposed by other 
authors~\cite{GLM,K1,K2}.
\subsubsection*{From HERA to the Tevatron}
The rate for diffractive $W$ production at the Tevatron 
can be calculated directly from 
$F_2^{D(3)}(Q^2,\beta,\xi)$~\cite{KGDIS,Whitmore}.
Using conventional factorization, the expected SD to ND ratio for $W$ 
production is 6.7\%~\cite{KGDIS}, while  
by scaling the normalization of the $1/\xi^{1+n}$ term in~(\ref{F2D3})
by the ratio of its integral at HERA ($\xi_{min}=Q^2/\beta s$) 
to its integral at the Tevatron ($\xi_{min}=M_0^2/\beta s$, with 
$M_0^2$=1.5 GeV$^2$) 
the prediction 
becomes 1.24\%, in agreement with the data.

\section*{CONCLUSION}
We have reviewed the pomeron flux renormalization hypothesis and 
compared expectations for renormalized soft 
and hard diffraction rates with available experimental results.
In all cases considered, soft and hard, 
the renormalized flux predictions are found to be in 
excellent agreement with the data. The renormalization procedure 
consists in simply scaling the standard pomeron flux to its integral 
over all available phase space. This integral is process dependent 
and therefore conventional factorization breaks down. Thus, the 
renormalization of the pomeron flux can be viewed as a scaling law in 
diffraction, which  unitarizes the diffractive amplitude at the expense of 
conventional factorization.\\

\end{document}